\documentclass[article,twocolumn]{revtex4-1}
\usepackage{graphicx}
\usepackage{color}
\newcommand{\hPi}{\hat{\Pi}}
\newcommand{\hR}{\hat{R}}

\newcommand{\had}{\hat{a}^+}
\newcommand{\hadp}{\hat{a}^{\prime +}}
\newcommand{\hbdp}{\hat{b}^{\prime +}}
\newcommand{\hadpp}{\hat{a}^{\prime \prime +}}
\newcommand{\hbdpp}{\hat{b}^{\prime\prime +}}
\newcommand{\ha}{\hat{a}}
\newcommand{\hbd}{\hat{b}^+}
\newcommand{\hcd}{\hat{c}^+}
\newcommand{\hb}{\hat{b}}
\newcommand{\hrho}{\hat{\rho}}
\newcommand{\vac}{|{\rm vac}\rangle}
\newcommand{\ket}[1]{\left| #1 \right\rangle}
\newcommand{\bra}[1]{\left\langle #1 \right|}
\newcommand{\braket}[2]{\left\langle #1 | #2 \right\rangle}
\newcommand{\proj}[1]{| #1\rangle\!\langle #1 |}
\newcommand{\Tr}{\mathrm{Tr}}

\newcommand{\eea}{\end{eqnarray}}
\newcommand{\bea}{\begin{eqnarray}}
\newcommand{\ee}{\end{equation}}
\newcommand{\be}{\begin{equation}}

\begin{document}
\title{The time-dependent spectrum of a single photon and its POVM}
\author{S.J. van Enk}
\affiliation{Department of Physics and
Oregon Center for Optical, Molecular \& Quantum Sciences\\
University of Oregon, Eugene, OR 97403}
\begin{abstract}
Suppose we measure the time-dependent spectrum of a single photon. That is, we first send the photon through a set of frequency filters (which we assume to have different filter frequencies but the same finite bandwidth $\Gamma$), and then record at what time (with some finite precision $\Delta t$, and with some finite efficiency $\eta$) and after passing what filter the photon is detected. What is the POVM  (Positive-Operator Valued Measure, the most general description of a quantum measurement) corresponding to such a measurement? We show how to construct the POVM in various cases, with special interest  in the case $\Gamma\Delta t\ll 1$ (time-frequency uncertainty still holds, even in that limit). One application of the formalism is to heralding single photons. We also find a Hong-Ou-Mandel type of interference effect with two photons entering a frequency filter.
\end{abstract}
\maketitle
\section{Introduction}

The time-dependent physical spectrum of light \cite{eberly1977} is an operationally defined quantity that characterizes a specific combination of spectral and temporal properties of any light signal.
The definition includes the use of a frequency filter with a finite bandwidth before the measurement of the light intensity at some time $t$. 

The measurement can be performed on a quantum light source as well, where instead of measuring intensity we count photons.
It turns out then that
the frequency filtering operation can play an active and nontrivial  role in the detection of two-photon correlations \cite{schrama1991,schrama1992,joosten2000}. In particular,
under certain conditions it may reverse the time order of two photons, in that the photon that was emitted earlier may be detected later.
In such a case, destructive interference between two opposite time orders of emitting a pair of photons may substantially reduce the joint probability of detecting both photons at the same time, in principle all the way down to zero \cite{schrama1991,schrama1992}. 
Moreover, the filtering process may change the photon statistics from antibunched to bunched \cite{joosten2000}.

Even for a single photon 
its time-dependent spectrum can reveal
interesting information, e.g., about its history.
For example,
when interacting with multiple different resonators, a photon could remain trapped in those resonators for a little while, but only when its frequency is close to one of the resonances. Thus, the photon's early time-dependent spectrum
may display dips at such resonance frequencies \cite{mirza2013}, with those frequency components emerging in the spectrum only at later times.
For another example, consider a single photon emitted from an optomechanical cavity that has one  moving mirror.
The spectrum will contain red sidebands (below the cavity resonance frequency) at early times, corresponding to the loss of photon energy to phonons, but at later times it can contain blue sidebands as well, as a result of the photon having gained energy from the moving mirror \cite{liao2012,mirza2014}.

We construct here the POVM (Positive-Operator Valued Measure) corresponding to the measurement of the single-photon time-dependent spectrum. The POVM is the most general sort of measurement allowed by quantum mechanics \cite{kraus1983,preskill1998}. One difference, in particular, compared to the usual notion of a hermitian operator as observable, is that different outcomes of a measurement do not necessarily correspond to projectors onto pure orthogonal states. We will see here that measuring the time-dependent spectrum of a single photon indeed involves projecting onto nonorthogonal single-photon states.
Moreover, the less precise the time measurement is, the larger is the effective Hilbert space dimension on which
one projects.

The operators thus obtained can be applied to the problem of heralding single photons by spectrally filtering and subsequently detecting one photon of a photon pair \cite{mcmillan2009}. 
This problem has recently gained interest, in particular because of the trade-offs between the purity of the heralded photon and the efficiency of the heralding process \cite{meyer2017,blay2017}. Here we will see that high purity of the heralded photon can be obtained with filtering and, at least in principle, without trading off for efficiency, provided the measurement performed after the filter is pure. 

We will also describe both single-photon and two-photon measurements with two different frequency filters,
as well as a two-photon Hong-Ou-Mandel type of interference effect \cite{hong1987} observable with one frequency filter.

\section{Spectral filtering}
\begin{figure}[htbp]
\begin{center}
\includegraphics[width=3in]{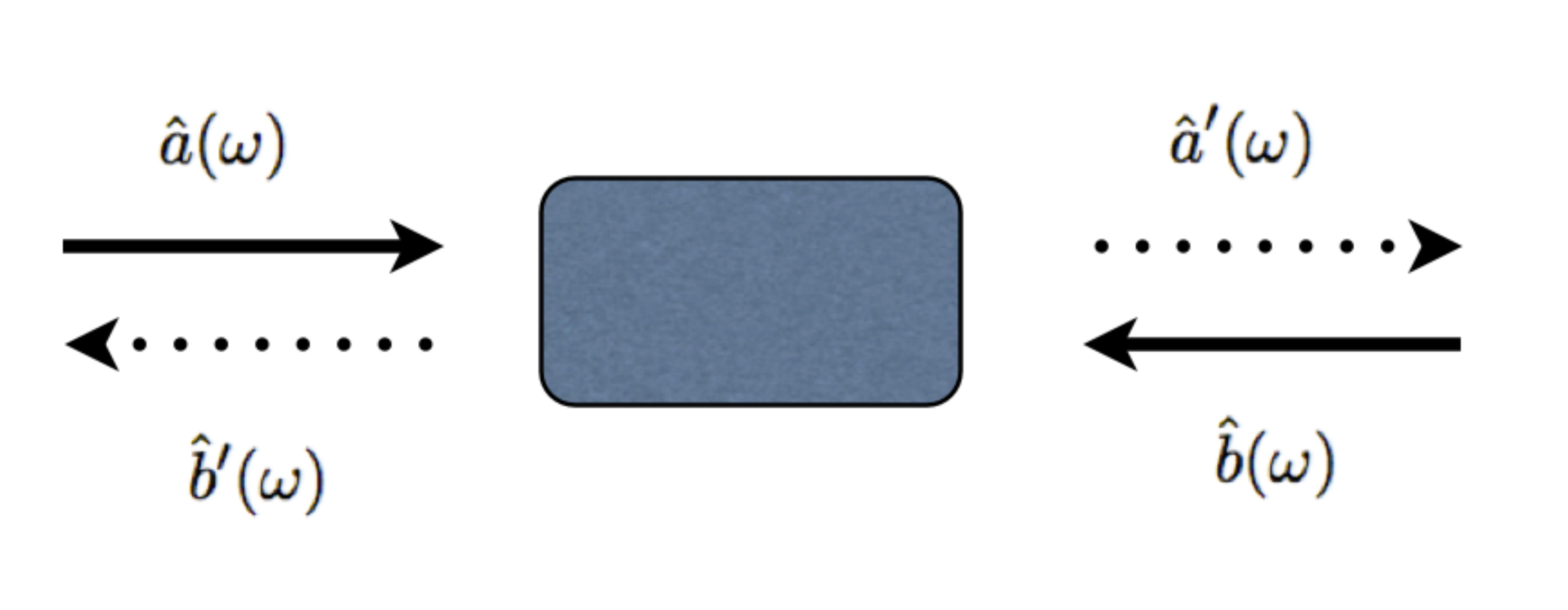}
\caption{Modes used to describe spectral filtering. There is one set of input modes of interest, which contains the light we are filtering, described by annihilation operators $\ha(\omega)$. There is a second set of auxiliary input modes, which are assumed to be in the vacuum state, described by $\hb(\omega)$. There are two sets of output modes: the operators $\ha'(\omega)$ describe light transmitted through the filter, and that is the light  we are going to perform measurements on with the aim to learn about the light in the input modes $\ha(\omega)$. The other ``reflected'' modes, described by $\hb'(\omega)$,  are available for further filtering operations (at different frequencies).}
\label{default}
\end{center}
\end{figure}

Spectral filtering by a passive (photon-number preserving and time-independent) device can be most conveniently described as a unitary transformation on two
sets of input modes (see Fig.~1) described by annihilation operators $\ha(\omega)$ and $\hb(\omega)$ and their hermitian conjugates, the creation operators, with $\omega>0$ the frequency of the modes   \cite{vogel2006},  
\bea\label{vogel}
\ha'(\omega)&=&T(\omega) \ha(\omega)
+R(\omega)\hb(\omega),\nonumber\\
\hb'(\omega)&=&R(\omega) \ha(\omega)
+T(\omega)\hb(\omega).
\eea
These creation and annihilation operators satisfy bosonic commutation relations, and in particular
$[\ha(\omega'),\had(\omega)]=\delta(\omega'-\omega)$. Operators acting on different modes commute.
Here one set comprises the input modes of interest, $\ha(\omega)$, which contain the light we are filtering, while the other set consists of auxiliary modes 
$\hb(\omega)$ that are assumed initially to be in the vacuum state. The two sets of modes $\ha'(\omega)$ and $\hb'(\omega)$ are output modes.
The functions $T(\omega)$ and $R(\omega)$
are complex transmission (through the spectral filter) and reflection (off the filter) coefficients
that satisfy
\bea\label{TRom}
|T(\omega)|^2+|R(\omega)|^2&=&1,\nonumber\\
T(\omega)R^*(\omega)+T^*(\omega)R(\omega)
&=&0.
\eea
The filter function $T(\omega)$ is peaked around a particular frequency of interest, $\omega_0$, and has a bandwidth $\Gamma$, which we may define as
\be\label{Gamma}
\Gamma=\frac{1}{\pi}\int_0^\infty d\omega\,
|T(\omega)|^2.
\ee
 We will always assume here that $\omega_0/\Gamma\gg 1$.

The modes on which we perform photo detection measurements are the transmitted output modes $\ha'(\omega)$, but the aim is to obtain information about the state of the input modes $\ha(\omega)$. 
Suppose we start with a pure single-photon input state
$\hrho=\proj{\phi}$ with
\be\label{one}
\ket{\phi}=\int_0^\infty\! d\omega\, \phi(\omega)\had(\omega)\vac,
\ee
with $\vac$ the vacuum state with no photons.
In order to describe how this state  is transformed by the filter we need the inverse transformation
of the creation operators, which we can obtain by making use of (\ref{TRom}):
\bea\label{TRo}
\had(\omega)&=&T(\omega) \hadp(\omega)
+R(\omega)\hbdp(\omega)\nonumber\\
\hbd(\omega)&=&R(\omega) \hadp(\omega)
+T(\omega)\hbdp(\omega),
\eea
which actually has the same form as the forward transformation (\ref{vogel}) of annihilation operators.
The output state can, therefore, be written as
\bea
\ket{\phi'}&=&\int_0^\infty\! d\omega\, 
T(\omega)\phi(\omega)\hadp(\omega)\vac\nonumber\\
&+&\int_0^\infty\! d\omega\, 
R(\omega)\phi(\omega)\hbdp(\omega)\vac,
\eea
where the first term describes a single photon in the transmitted output modes and  zero photons in the 
reflected modes, whereas for the second term the one single photon is in the reflected modes.

Further information about the same input state can be obtained by having a second filter, with a different filter frequency $\omega_1$ (for convenience we assume all filters to have the same bandwidth $\Gamma$).
That filter should be applied to the reflected modes $\hb'(\omega)$, and the second measurement will then be done on the modes $\ha''(\omega)$ transmitted through that second filter. 
The output state (of three different sets of output modes)  will in this case be
\bea\label{TR}
\ket{\phi''}&=&\int_0^\infty\! d\omega\, 
T(\omega)\phi(\omega)\hadp(\omega)\vac\nonumber\\
&+&\int_0^\infty\! d\omega\, 
T_1(\omega)R(\omega)\phi(\omega)\hadpp(\omega)\vac\nonumber\\
&+&\int_0^\infty\! d\omega\, 
R_1(\omega)R(\omega)\phi(\omega)\hbdpp(\omega)\vac.
\eea
With $N>2$ filters, photo detection measurements are to be performed on $N$ transmitted output modes, with one reflected set of modes available for yet another filter. 
The generalization of (\ref{TR}) to more than two filters should be obvious.

Finally, a simple model expression  for $T(\omega)$ (valid only for frequencies close to $\omega_0$) is  \cite{vogel2006}
\be\label{modelT}
T(\omega)=\frac{\Gamma}{\Gamma-i(\omega-\omega_0)}.
\ee
We'll sometimes use this expression for illustrative purposes.
In that case we can take the corresponding reflection coefficient to be
\be\label{modelR}
R(\omega)=\frac{-i(\omega-\omega_0)}{\Gamma-i(\omega-\omega_0)}=1-T(\omega).
\ee

\section{An ideal measurement}
As a warm-up exercise imagine first we perform an idealized single-photon measurement on just the output of the first filter, described by 1D projectors $\proj{\phi_k}$, for $k=1\ldots\infty$. Here $\ket{\phi_k}$ is a pure single-photon state defined as 
\be
\ket{\phi_k}=\int_0^\infty\! d\omega\, \phi_k(\omega)\hadp(\omega)\vac,
\ee
 in terms of an orthonormal set of mode functions satisfying
\be
\int_0^\infty \!d\omega\, \phi_k^*(\omega)\phi_{k'}(\omega)=\delta_{kk'}.
\ee
The probability to detect the photon in the output  and get outcome $k$ is 
\be\label{Pk}
P_k=\left|\int_0^\infty d\omega\,\phi_k^*(\omega) T(\omega)\phi(\omega)\right|^2.
\ee
This probability can be rewritten in terms of a POVM element
$\hPi_k$ as $P_k=\Tr(\hrho\hPi_k)$, provided we choose
\be
\hPi_k=w_k\proj{T\phi_k}.
\ee
Here
$w_k$ is a weight (nonnegative and $\leq 1$)
\be
w_k=\int_0^\infty\! d\omega\, |T(\omega)|^2|\phi_k(\omega)|^2,
\ee
and $\ket{T\phi_k}$ is a pure and properly normalized {\em input} single-photon state 
\be\label{Tphi}
\ket{T\phi_k}=\frac{1}{\sqrt{w_k}}\int_0^\infty\! d\omega \, 
T^*(\omega) \phi_k(\omega)
\had(\omega)\vac.
\ee
Note the appearance of $T^*$ here, rather than $T$. Indeed, the state $\ket{T\phi_k}$ does not describe a photon that initially is in the state $\ket{\phi_k}$ and is then frequency filtered, but, rather, a state the photon could have been in before the filter, if it is later detected in the state $\ket{\phi_k}$.

In fact, the state (\ref{Tphi}) can also be obtained
by starting with the state $\ket{\phi_k}$ at the output and propagating it back in time (using the inverse of (\ref{TRo})). This yields a (normalized) state  containing one photon spread out over  {\em both} input modes
$\ha(\omega)$ and $\hb(\omega)$. Subsequently projecting onto the vacuum state of the input modes $\hb(\omega)$ leaves one with the subnormalized state $\sqrt{w_k}\ket{T\phi_k}$ for just the input modes $\ha(\omega)$. This procedure of propagating backwards in time from the final signal to the input may give  the clearest explanation of what a POVM actually is.

The physical meaning of $w_k$ is as follows: even if the input photon is in the state $\ket{T\phi_k}$, outcome $k$ is not guaranteed. The probability of getting that measurement outcome is $w_k$. There are two reasons
for not getting outcome $k$ with 100\% probability. The first is that
the projectors for different values of $k$ are nonorthogonal because of the presence of the filter function:
\be
\braket{T\phi_k}{T\phi_{k'}}=\frac{1}{\sqrt{w_kw_{k'}}}\int_0^\infty\! d\omega\,
|T(\omega)|^2\phi_k^*(\omega)\phi_{k'}(\omega).
\ee
And so we may obtain outcome $k'\neq k$
for input state $\ket{T\phi_k}$. Secondly,
 we may not actually detect the photon at all, because the input photon may have been reflected off of the frequency filter.
One way to express the probability of a null detection on the transmitted mode is by equating it with the probability we would detect the reflected photon. Imagining we perform the same measurement $\proj{\phi_k}$ on the reflected mode
$\hb'(\omega)$ we would get outcome $k$ with probability
\be
Q_k=\left|\int_0^\infty\! d\omega\,\phi_k^*(\omega) R(\omega)\phi(\omega)\right|^2,
\ee
where we just replaced $T$ with $R$ in (\ref{Pk}).
And just as before, this probability $Q_k$ can be rewritten in terms of a POVM element
$\hR_k$ as $Q_k=\Tr(\hrho\hR_k)$, provided we pick
\be
\hR_k=u_k\proj{R\phi_k},
\ee
where the weight
$u_k$ is
\be
u_k=\int_0^\infty\! d\omega\, |R(\omega)|^2|\phi_k(\omega)|^2=1-w_k,
\ee
and the single-photon state
$\ket{R\phi_k}$ is 
\be
\ket{R\phi_k}=\frac{1}{\sqrt{u_k}}\int_0^\infty\! d\omega \, 
R^*(\omega) \phi_k(\omega)
\had(\omega)\vac.
\ee
We can now write the POVM element for {\em not} detecting a photon in the output modes of interest, $\ha'(\omega)$, as
\be\label{null}
\hPi_{{\rm null}}=\sum_k u_k\proj{R\phi_k}.
\ee
The complete POVM corresponding to the ideal single-photon measurement on $\ha'(\omega)$ is then
\be
\{(\hPi_k)_{k=1}^{\infty},\hPi_{{\rm null}}\},
\ee
which indeed obeys the relation
\be
\hPi_{{\rm null}}+\sum_{k=1}^\infty \hPi_k =\openone_1,
\ee
where $\openone_1$ is the identity  on the single-photon subspace of the input modes $\ha(\omega)$.
For, if we define
the Dirac-normalized ket $\ket{\omega}= \had(\omega)\vac$, then we have
\bea
\hPi_{{\rm null}}&=&\int_0^\infty\! d\omega\, |R(\omega)|^2
\proj{\omega},\\
\sum_{k=1}^\infty \hPi_k &=&\int_0^\infty\!  d\omega\, |T(\omega)|^2
\proj{\omega},\label{POVM}\\
\openone_1&=&\int_0^\infty\!  d\omega \,
\proj{\omega}.
\eea
Even though the actual photo detection measurement is performed on the output modes $\ha'(\omega)$ the POVM
refers to the input modes $\ha(\omega)$.
\subsection{Heralding a single photon}\label{herald}
It is not straightforward to create a single photon. One technique involves downconversion, a nonlinear optical process in which one photon of high frequency is (with some (very) small probability) converted into two photons of lower frequency. The state produced consists of a large vacuum component plus a small two-photon component (as well as an even smaller four-photon component, etc.). By detecting one of the two photons we project out the vacuum component, so that the remaining state has a large single-photon component.

The two-photon part of the state is to a good approximation pure, and we may write
\bea
\ket{\Phi}=\int_0^\infty \!\!d\omega
\int_0^\infty\!\! d\omega' \,\Phi(\omega,\omega')
\had(\omega)\hcd(\omega')\vac,
\eea
where we normalize 
\be
\int_0^\infty \!\!d\omega
\int_0^\infty\!\! d\omega' \,|\Phi(\omega,\omega')|^2=1,
\ee
such that $\braket{\Phi}{\Phi}=1$.
We assume we detect the (``heralding'') photon in modes
$\had(\omega)$ and then infer the presence of the heralded photon in modes $\hcd(\omega')$.
We would prefer the heralded photon to be pure.
High purity of the heralded photon can be achieved {\em without} filtering by engineering the function $\Phi(\omega,\omega')$ \cite{grice2001,mosley2008}, or with filtering \cite{mcmillan2009}. In the latter case, however, there will in practice be a trade-off between purity and the efficiency of the heralding process, i.e., the probability to detect the heralding photon \cite{meyer2017,blay2017}.

Here, where we assume that an {\em ideal} measurement is performed after filtering, the heralded photon is actually always pure. We detect the heralding photon and obtain outcome $k$ with probability
\be
{\cal P}_k=\int_0^\infty \!\!d\omega'\,
\left|\int_0^\infty\!\! d\omega \,T(\omega)\phi_k^*(\omega)\Phi(\omega,\omega')\right|^2.
\ee
The density matrix of the heralded photon conditioned on outcome $k$ may be written as
\be
\hrho_{c|k}=\frac{1}{{\cal P}_k}
\Tr_a \left(\sqrt{\hPi_k}\proj{\Phi}\sqrt{\hPi_k}\right),
\ee
where the trace is taken over the $\ha(\omega)$ modes. With $\hPi_k$ projecting onto a pure state, $\hrho_c^{(k)}$ corresponds to a pure state, too:
\be
\ket{\psi_{c|k}}=
\frac{\int_0^\infty \!\!d\omega
\int_0^\infty\!\! d\omega' T(\omega)\phi_k^*(\omega)\Phi(\omega,\omega')\hcd(\omega')\vac}{\sqrt{{\cal P}_k}}.
\ee
On the other hand, if we ignore the information about which outcome we obtained or if our measurement does not distinguish between different outcomes $k$, we get a mixed state
\be\label{rhoc}
\hrho_c=\frac{1}{{\cal P}}
\sum_k\Tr_a (\proj{\Phi}\hPi_k)
\ee
 where
\be
{\cal P}=\sum_k {\cal P}_k.
\ee
Here we made use of the cyclical property of the trace to move one $\sqrt{\hPi}$ factor next to the other.
We may now substitute (\ref{POVM})
for $\sum_k \hPi_k$ into (\ref{rhoc}) to get an expression equivalent to that obtained in Ref.~\cite{blay2017}:
\be
\hrho_c=\frac{1}{{\cal P}}\int_0^\infty\!  d\omega\,
|T(\omega)|^2
\bra{\omega}\proj{\Phi}\ket{\omega} 
\ee
This expression is useful because it provides a lower bound on the purity of the heralded photon (the more information we have about which outcome occurred, the higher the purity). In fact, it is this lower bound that displays the trade-off between purity and efficiency: by making the filter narrower we would increase the purity of the heralded photon but at the cost of decreasing the probability ${\cal P}$ of detecting the heralding photon in the first place. 
\section{Measuring the time-dependent spectrum}
Suppose we do not perform an ideal single-photon measurement, but instead merely detect the presence of the photon (in the frequency-filtered output modes) within a finite time interval $I$ between $t_0$ and $t_0+\Delta t$.  
Whereas measuring the frequency of light has been relatively straightforward since Newton's days, measuring the time of arrival (on, say, a picosecond timescale) tends to be much harder. One nice way to achieve a high-resolution time measurement is to measure the frequency after first sending the photon through a time-to-frequency converter \cite{beck1993,kauffman1994,wong1994,bennett1999,azana2003,azana2004}. Here  one first lets the photon propagate through a dispersive element that multiplies the spectral amplitude of the photon with a phase factor $\exp(-i\alpha \omega^2/2)$, with $\alpha$ a constant, and subsequently one applies a time-dependent phase modulation that multiplies the temporal amplitude with a similar phase shift $\exp(-i\beta t^2/2)$ in the time-domain. For the special choice $\beta=1/\alpha$ the combined transformation applied to the photon is actually the Fourier transform, transforming time to frequency (the mathematics involved is exactly analogous to that describing spatial light propagation through a cylindrical lens, hence the name: ``time lens'').
The time resolution thus obtainable extends down to hundreds of femtoseconds \cite{bennett1999,azana2004}.

\subsection{One photon and a single frequency filter}\label{single}

Consider a photon in a pure input state $\hrho=\proj{\phi}$. The probability to detect that photon  in the filtered output modes between times $t_0$ and $t_0+\Delta t$ would be
\be
P_I=\frac{1}{2\pi}\int_{t_0}^{t_0+\Delta t}\!\!\!\!dt\,\, \left|\int_0^\infty \! d\omega\, T(\omega)\phi(\omega){\rm e}^{-i\omega t}\right|^2,
\ee
if we had a perfectly efficient photodetector.
If the detector has a frequency-dependent efficiency $\eta(\omega)$, 
we have
\be\label{PI}
P_I=\frac{1}{2\pi}\int_{t_0}^{t_0+\Delta t}\!\!\!\!dt\,\, \left|\int_0^\infty \! d\omega\, \sqrt{\eta(\omega)}T(\omega)\phi(\omega){\rm e}^{-i\omega t}\right|^2.
\ee
In the following we assume for simplicity that $\eta(\omega)$ does not depend on $\omega$ over the relevant frequency range(s) and so we can always take it out of integrals over $\omega$. We'll denote the constant efficiency by $\eta$.

We can write the probability (\ref{PI}) as $P_I=\Tr(\hrho\hPi_I)$ with $\hPi_I$ an integral over $t$-dependent POVM elements as
\be\label{hPiI}
\hPi_I=\int_{t_0}^{t_0+\Delta t}\!\!\! dt
\, w\proj{\Psi_t}, 
\ee
with the normalized single-photon state
$\ket{\Psi_t}$ defined as
\be
\ket{\Psi_t}=\frac{\sqrt{\eta}}{\sqrt{2\pi w}}\int_0^\infty\! d\omega\,
T^*(\omega){\rm e}^{i\omega t} \had(\omega)\vac,
\ee
and the weight density (per unit of time) $w$ is
\be
w=\frac{\eta}{2\pi}\int _0^\infty\! d\omega\,
|T(\omega)|^2=\eta\Gamma/2,
\ee
where the second equality follows from our definition (\ref{Gamma}) of the bandwidth.
Note that the time parameter in $\ket{\Psi_t}$ refers to the time of detection, not the evolution in time of the input state.

The POVM elements corresponding to detecting the photon at different times are not orthogonal, and we have
\bea\label{Psitt}
\braket{\Psi_{t}}{\Psi_{t'}}=
\frac{1}{\pi\Gamma}\int_0^\infty \!d\omega\,
|T(\omega)|^2{\rm e}^{i\omega(t'-t)}\nonumber\\
\approx
{\rm e}^{-\Gamma |t'-t|}{\rm e}^{i\omega_0(t'-t)},
\eea
where the approximate equality is valid for the model expression (\ref{modelT})
for $T(\omega)$, under the assumption that $\omega_0\Gamma\gg 1$. The latter assumption allows one to extend the integral over $\omega$ to negative frequencies, after which the integral can be performed by contour integration in the complex plane.

If we add up all POVM elements corresponding to detecting the photon in a complete set of non-overlapping time intervals we obtain the same result (\ref{POVM}). To see this, we could take the limits  $t_0\rightarrow -\infty$ and $\Delta t\rightarrow\infty$
for the single POVM element (\ref{hPiI}).

The POVM element $\hPi_I$ is not simply one projector onto a pure state but a mixture of such projectors. That is, $\hPi_I$ is not {\em pure}.
We can define the purity of an arbitrary POVM element $\hPi$ in analogy to the definition of purity of a quantum state $\hrho$, namely $\Tr(\hrho^2)$, as
\be
{\rm Pur}(\hPi)=\frac{\Tr(\hPi^2)}{(\Tr(\hPi))^2},
\ee
where the denominator accounts for the fact that $\Tr(\hPi)$ does not have to equal 1, whereas $\Tr(\hrho)=1$ for any physical state. 
The definition is such that
\be
0\leq {\rm Pur}(\hPi)\leq 1,
\ee
with ${\rm Pur}(\hPi)=1$ only for
$\hPi$ proportional to a 1D projector when $\hPi$ can be written as $\hPi=w \proj{\psi}$.
The lower limit of the purity is actually $1/d$ with $d$ the dimension of the Hilbert space on which $\hPi$ acts, but here $d$ is infinite.
Given a value for the purity we can conversely define an effective Hilbert subspace dimension on which $\hPi$ projects by
\be
d_{{\rm eff}}=1/{\rm Pur}(\hPi).
\ee
This effective dimension has the meaning of the number of different (orthogonal) input states that lead to the same measurement outcome.
For $\hPi_I$ we have
\bea
\Tr(\hPi_I)&=&\frac{\eta\Gamma\Delta t}{2}\nonumber\\
\Tr(\hPi_I^2)&=&\frac{\eta^2\Gamma^2}{4}\int_0^{\Delta t}\!dt\,
\int_0^{\Delta t}\!dt'\,|\braket{\Psi_{t}}{\Psi_{t'}}|^2\nonumber\\
&\approx& \frac{\eta^2}{8}\left[
{\rm e}^{-2\Gamma\Delta t}+2\Gamma\Delta t -1\right],
\eea
with the last approximate equality 
following from the last line of (\ref{Psitt}).
For that case, the purity of $\hPi_I$ is approximately (note the efficiency $\eta$ drops out)
\be
{\rm Pur}(\hPi_I)\approx
\frac{{\rm e}^{-2\Gamma\Delta t}+2\Gamma\Delta t-1}{2(\Gamma\Delta t)^2}.
\ee 
In the limit of large $\Gamma\Delta t$ the purity goes to zero as $1/(\Gamma\Delta t)$,  which signifies loss of information, from having performed a timing measurement that is less precise than was in principle possible. The effective Hilbert space dimension is $d_{{\rm eff}}\approx\Gamma\Delta t+1/2$ in this case.

Measurements with $\Gamma\Delta t\ll 1$, on the other hand, are the most interesting.
In that case the POVM element $\hPi_I$ becomes pure, i.e., ${\rm Pur}(\hPi_I)\rightarrow 1$, and we can approximate
\be\label{hP}
\hPi_I\approx \frac{\eta\Gamma\Delta t}{2}
\proj{\Psi_{t_0+\Delta t/2}}.
\ee
This means that when $\Delta t$ is short (compared to $\Gamma^{-1}$), the measurement projects onto a pure single-photon state with a central frequency of $\omega_0$
and a width in time not determined by the short detection time interval $\Delta t$, but by the much longer filtering time $\Gamma^{-1}$.
This explains why two photons  that are less than a distance $\Gamma^{-1}$ apart in time may well be detected in the opposite time order,
when they pass through a narrow bandwidth filter first (see also Section \ref{two}). 

In the limit $\Gamma\Delta t\rightarrow 0$ the POVM would be pure and so such a POVM used for heralding a single photon would yield a pure heralded photon, just as in Section \ref{herald}. In the opposite limit
one reaches the result (\ref{rhoc}). For any finite value of $\Gamma\Delta t$ the result lies somewhere between these two extreme results. It is an open question to what extent the purity vs efficiency trade-off relation can be softened in this case \cite{blay2017} \footnote{Daniel R. Blay, private communication.}.

\subsection{One photon and two frequency filters}
It is straightforward now to take into account 
the possibility of filtering multiple different colors (but still assuming one single photon in the input). With a second filter in place, described by a resonance frequency $\omega_1$ and a bandwidth $\Gamma$,
we need the corresponding transmission coefficient $T_1(\omega)$. Since that filter should be placed in the beam that is reflected off of the first filter, we also need the reflection coefficient $R(\omega)$ of the first filter (cf. Eq.~(\ref{TR})). The probability of detecting the photon during the time interval $I$ after the second filter, therefore, is
\be
P'_I=\frac{\eta}{2\pi}\int_{t_0}^{t_0+\Delta t}\!\!\!\!dt\,\, \left|\int_0^\infty \! d\omega\, T_1(\omega)R(\omega)\phi(\omega){\rm e}^{-i\omega t}\right|^2.
\ee
And so this measurement is described by the POVM element
\be
\hPi'_I=\int_{t_0}^{t_0+\Delta t}\!\!\! dt
\, w'\proj{\Psi'_t}, 
\ee
where the form of the normalized single-photon state is now slightly different than that corresponding to just one filter because of the presence of an additional factor $R^*$ (and, of course, with $T_1$ replacing $T$):
\be
\ket{\Psi'_t}=\frac{\sqrt{\eta}}{\sqrt{2\pi w'}}\int_0^\infty\! d\omega\,
R^*(\omega)T_1^*(\omega){\rm e}^{i\omega t} \had(\omega)\vac.
\ee
The weight density $w'$ is likewise
a bit more complicated than with just one filter:
\be
w'=\frac{\eta}{2\pi}\int _0^\infty\! d\omega\,
|R(\omega)T_1(\omega)|^2.
\ee
We may note that $w'\leq w$.
In the case of two very different filters, with $|\omega_1-\omega_0|\gg \Gamma$, we can replace $R(\omega)$ by 
1 in all the above integrals and obtain results that have the same form as for one filter, with $T_1$ replacing $T$, and hence with $\omega_1$ replacing $\omega_0$. (In that limit, $w'\rightarrow w$.)

Two POVM elements $\proj{\Psi'_t}$ and $\proj{\Psi_t}$ corresponding to detecting at two different frequencies (but at the same time $t$) are nonorthogonal, and become orthogonal only in the limit
$\Gamma/|\omega_1-\omega_0|\rightarrow 0$. For two different measurement outcomes corresponding to different photo detection times {\em and} different filter frequencies, we have 
\bea
\braket{\Psi'_{t'}}{\Psi_{t}}=
\frac{\int_0^\infty \!d\omega\,
T^*(\omega)R(\omega)T_1(\omega){\rm e}^{i\omega(t-t')}}{2\pi\sqrt{ww'}/\eta}.
\eea

\subsection{Two photons and two frequency filters}\label{two}
In the experiments reported in Ref.~\cite{schrama1992} {\em two} photons of different color are detected with the help of two different frequency filters. We can describe that situation with the help of a POVM, too.
Detection at time $t'$ of a photon after the second filter {\em and} a photon at time $t$ after the first filter
projects back onto a two-photon state of the input modes $\ha(\omega)$---again making use of the fact that the modes $\hb(\omega)$ start off in the vacuum state---of the form
\begin{widetext}
\be
\ket{\Psi_{t,t'}}=\frac{\eta}{2\pi\sqrt{W}}
\int_0^\infty\!d\omega\,
\int_0^\infty\!d\omega'\,
T^*(\omega)
T_1^*(\omega')R^*(\omega'){\rm e}^{i\omega t}{\rm e}^{i\omega' t'}\had(\omega)\had(\omega')\vac.
\ee
\end{widetext}
We can write $W$ as a sum of  two contributions here,
\bea\label{W}
W&=&\frac{\eta^2}{4\pi^2}\int_0^\infty\!d\omega\,
\int_0^\infty\!d\omega'\,
|T(\omega)|^2|R(\omega')T_1(\omega')|^2+\nonumber\\
&&
\frac{\eta^2}{4\pi^2}\left|
\int_0^\infty\!d\omega\,
T^*(\omega)T_1(\omega)R(\omega){\rm e}^{i\omega(t-t')}\right|^2.\nonumber\\
\eea
In fact, we can write this sum also in the form
\be\label{W2}
W=ww'\left(1+|\braket{\Psi'_{t'}}{\Psi_{t}}|^2\right),
\ee
thus making a connection between the one two-photon measurement and two single-photon measurements (on separate single-photon input states). 

The two terms in $W$ [either in (\ref{W}) or in (\ref{W2})] can be interpreted as follows. If the input state is actually
$\ket{\Psi_{t,t'}}$, then we obtain the measurement outcome with probability $W$.
The first term of $W$ is a product of two probabilities: the probability that a photon in the input state
\be\label{1}
\ket{T_t}\propto\int_0^\infty\!d\omega\,
T^*(\omega)
{\rm e}^{i\omega t}\had(\omega)\vac
\ee
is detected after the first filter at time $t$, and
 the probability that a photon in the input state
\be\label{2}
\ket{T_1R_{t'}}\propto\int_0^\infty\!d\omega'\,
T_1^*(\omega')R^*(\omega'){\rm e}^{i\omega' t'}\had(\omega')\vac
\ee
is detected after the second filter at time $t'$.

The second term  is also a product of two probabilities:  the probability that the photon in state (\ref{1})
is detected after the {\em second} filter at time $t'$ and the (equal!) probability that the photon in state (\ref{2}) is detected after the {\em first} filter at time $t$.
This second combination of two detection events is
likely only if both detection times are close to each other (within $\sim\Gamma^{-1}$) and if both filter frequencies are close, too (within $\sim\Gamma$).

\subsection{Two-photon interference with a frequency filter}
So far we have assumed the input modes $\hb(\omega)$ are in the vacuum state. But if we allow an input state that has one photon in each of the two input modes, such that
\be
\ket{{\rm in}}=\int_0^\infty\!d\omega\,\int_0^\infty\!d\omega'\,
\phi_1(\omega)\phi_2(\omega')\had(\omega)
\hbd(\omega')\vac,
\ee
(so that the two photons are independent, i.e., neither correlated nor entangled)
interference effects of the Hong-Ou-Mandel type \cite{hong1987} become possible.
In particular, with the help of (\ref{TRo}), the output state after the frequency filter can be written as
\begin{widetext}
\bea
\ket{{\rm out}}&=&\int_0^\infty\!d\omega\,\int_0^\infty\!d\omega'\,
\phi_1(\omega)\phi_2(\omega')[T(\omega) \hadp(\omega)
+R(\omega)\hbdp(\omega)]
[R(\omega') \hadp(\omega')
+T(\omega')\hbdp(\omega')]\vac\nonumber\\
&=&
\left[\int_0^\infty\!d\omega\,\int_0^\infty\!d\omega'\,
\phi_1(\omega)\phi_2(\omega')
T(\omega)R(\omega') \hadp(\omega) \hadp(\omega')+\right.\nonumber\\
&&\int_0^\infty\!d\omega\,\int_0^\infty\!d\omega'\,
\phi_1(\omega)\phi_2(\omega')R(\omega)T(\omega')\hbdp(\omega)
\hbdp(\omega')+\nonumber\\
&&\left.
\int_0^\infty\!d\omega\,\int_0^\infty\!d\omega'\,
\huge[\phi_1(\omega)\phi_2(\omega')
T(\omega)T(\omega') +
\phi_1(\omega')\phi_2(\omega)
R(\omega')R(\omega)\huge]\hadp(\omega)
\hbdp(\omega')\right]\vac.
\eea
\end{widetext}
It is the last line that contains interference effects. Namely, the two terms in the last line both describe an output state with one photon in the $\ha'$ modes and one photon in the $\hb'$ modes, but the terms cancel each other for every pair of frequencies $\omega,\omega'$ such that
\be\label{des}
\frac{\phi_1(\omega)
T(\omega)}{\phi_2(\omega)R(\omega)}=-
\frac{\phi_1(\omega')
R(\omega')}{\phi_2(\omega')T(\omega')}.
\ee
That is, when this condition is fulfilled we cannot detect a photon with frequency $\omega$ in output modes $\ha'$ if we detect one with frequency $\omega'$ in modes $\hb'$.
This type of interference is insensitive to phase shifts in the initial state, in that multiplying $\phi_{1,2}(\omega)$ by any frequency-independent phase factors does not impact the condition (\ref{des}) for destructive interference.
For $\omega=\omega'$ we get destructive interference if and only if
\be
|T(\omega)|^2=|R(\omega)|^2=1/2,
\ee
the derivation of which makes use of both unitarity conditions (\ref{TRom}).
That is, at those frequencies $\omega$ for which
the transmission probability is exactly 1/2, 
we can only detect both photons in the same output mode, either both in the $\hadp(\omega)$ modes or (with the same probability 1/2)  in the $\hbdp(\omega)$ modes. This destructive interference is very similar to the original Hong-Ou-Mandel effect, but Eq.~(\ref{des}) describes a more general form (involving two colors \cite{raymer2010}) of the effect.

\section{Conclusions}
We showed how to construct the POVM corresponding to various measurements on single photons or on pairs of photons that involve one or more frequency filters.
Whereas the measurements are performed on light that exits the frequency filter(s),  the POVM describes on what states of the input modes that enter the filter(s) the measurement projects. These states in general are not orthogonal for different measurement outcomes. 

The POVM element describing the time-dependent spectrum of a single photon
projects onto single-photon states with a width in time set by the filtering time $\Gamma^{-1}$, not by the accuracy $\Delta t$ of the time of detection after the frequency filter. This is such that the standard time-frequency uncertainty relation is automatically satisfied, even  in the limit  $\Gamma\Delta t\ll 1$. In that limit  the POVM becomes {\em pure} (each element involving a projection onto a single pure state). In the opposite limit of large $\Gamma\Delta t$ each POVM element projects onto a Hilbert subspace with effective dimension $d_{{\rm eff}}\approx \Gamma\Delta t+1/2$. All these conclusions are independent of the finite efficiency of the photodetector(s) used.

As an application of the formalism we considered the process of heralding one photon by detecting another photon
after a frequency filter \cite{mcmillan2009,meyer2017,blay2017}. 
The approach used here leads to identifying best-case ($\Gamma\Delta t\rightarrow 0$) and worst-case ($\Gamma \Delta t\rightarrow\infty$) scenarios for the purity of the heralded photon, and all cases in between.
It is an open question how this exactly affects the purity-efficiency tradoff relation \cite{meyer2017,blay2017} for the heralding process.

Finally, we also found a Hong-Ou-Mandel type of interference effect between two photons (which could have  different colors) entering the two different input ports of a frequency filter.

\bibliography{spectral_POVM3}
\end{document}